\documentclass[onecolumn]{revtex4}

\usepackage{graphicx}
\usepackage{dcolumn}
\usepackage{color}
\usepackage{bm}
\usepackage{amsmath}
\usepackage{amsfonts}
\usepackage{amssymb}

\input epsf

\begin{document}

\title{A Study of Universal Thermodynamics in Brane World Scenario}

\author{Saugata Mitra\footnote{saugatamitra20@gmail.com}}
\affiliation{Department of Mathematics, Jadavpur University,\\
 Kolkata-700032, West Bengal, India.}
\author{Subhajit Saha\footnote {subhajit1729@gmail.com}}
\affiliation{Department of Mathematics, Jadavpur University,\\
 Kolkata-700032, West Bengal, India.}
\author{Subenoy Chakraborty\footnote {schakraborty.math@gmail.com}}
\affiliation{Department of Mathematics, Jadavpur University,\\
 Kolkata-700032, West Bengal, India.}

\begin{abstract}

A study of Universal thermodynamics is done in the frame work of RSII brane model and DGP brane scenario.
The Universe is chosen as FRW model bounded by apparent or event horizon. Assuming extended Hawking temperature on the horizon, the unified first law is examined for perfect fluid (with constant equation of state) and modified Chaplygin gas model. As a result there is a modification of Bekenstein entropy on the horizons. Further the validity of the generalized second law of thermodynamics and thermodynamical equilibrium are also investigated.

\end{abstract}

\maketitle

From astrophysical observations \cite{r1,r2,r3,r4,r5}, it is now well established that our Universe is going through an accelerating phase. It is speculated that
this cosmic acceleration is driven by some invisible fluid (known as dark energy (DE)) having strong repulsive gravitational effect and has come into
 action only in recent past. But till now the nature of DE is completely unknown to us and is an unresolved problem in modern theoretical physics
\cite{r6,r7,r8,r9,r10,r11,r12}. On the other hand people have tried to modify Einstein's gravity theory itself as an alternative way of resolving this problem. Brane world scenario is one of such
models related to gravity theory in higher dimensions. In RSII brane model (\cite{r13,r14,r15}), our Universe is a positive tension 3-brane embedded in a 5-dimensional
AdS bulk space-time. The standard model fields are confined on the brane while gravity can propagate in the bulk also. So the effective gravity on the
brane is different from the standard Einstein gravity due to the existence of extra dimension.
Another simple and well studied model of brane gravity is the Dvali-Gabadadze-Porrati (DGP) brane world model \cite{r16,r17,r18,r19}. In contrast to RSII model where
the extra dimension is of finite size, in DGP brane model our 4-dimensional world (3-brane) is embedded in a space-time with an infinite size
extra dimension with the motivation of resolving the cosmological constant problem as well as problems in  supersymmetry breaking \cite{r16,r17,r18,r19}. Usually in
this model, FRW brane is embedded in a 5-dimensional Minkowski bulk.

Black hole thermodynamics \cite{r20,r21,r22} and AdS/CFT correspondence \cite{r23}, established a deep connection between gravity and thermodynamics. Jacobson \cite{r24} and Padmanabhan \cite{r25,r26,r27} also showed the connection between gravity and thermodynamics. Jacobson deduced the Einstein's field equations from the Clausius relation for local Rindler horizon. Padmanabhan, on the other hand, showed that the field equations in Einstein gravity for a spherically symmetric space-time can be expressed as the first law of thermodynamics.

From thermodynamical viewpoint, the study of dynamical black hole was initiated by Hayward \cite{r28,r29,r30,r31}. He introduced the notion of
 trapping horizon in 4D Einstein gravity for non-stationary spherically symmetric spacetimes and showed that Einstein's equations are equivalent
 to the unified first law. Then projecting the unified first law along any tangential direction $ (\xi) $ to the trapping horizon, one is able
to derive the first law of thermodynamics \cite{r32,r33,r34} or equivalently Clausius relation of the dynamical black hole.
The homogeneous and isotropic FRW Universe may be considered as dynamical spherically symmetric space-time from cosmological viewpoint.

Further, our Universe is considered as a non-stationary gravitational system in the perspective of Universal thermodynamics. The inner trapping horizon coincides with the apparent horizon and one can study Universal thermodynamics using the unified first law. Starting with the unified first law, the Friedmann equations with arbitrary spatial curvature were derived by Cai and Kim \cite{r35}. They have considered   $T=\frac{1}{2\pi R_A }$ as the Hawking temperature and $S=\frac{ \pi R_{A}^2 }{G}$ as the Bekenstein entropy on the apparent horizon having radius $R_A$.
Also using the entropy formulae (not the Bekenstein one) for the static spherically symmetric black hole horizons in Gauss-Bonnet gravity and in more general Lovelock gravity they were also able to obtain the Friedmann equations in those gravity theories. Subsequently, Cai and Cao \cite{r32} have shown that Clausius relation do not hold at the apparent horizon of the FRW Universe in scalar-tensor gravity and they concluded that it corresponds to a system of non-equilibrium thermodynamics similar to f(R)-gravity (Eling et al. \cite{r36}). Cai and Cao \cite{r34} also studied thermodynamics of apparent horizon in RSII brane scenario and have obtained non-Bekenstein entropy on the horizon from the Clausius relation and it reduces to Bekenstein entropy in the limit of large horizon radius.

In the present work, we study Universal thermodynamics both in RSII brane model and DGP brane scenario using extended Hawking temperature on the Horizon (apparent / event) and investigate whether the entropy on the horizon is Bekenstein or not for the validity of the unified first law of thermodynamics. Finally, validity of the generalized second law of thermodynamics (GSLT) and thermodynamical equilibrium (TE) are also examined.

We start with homogeneous and isotropic FRW metric as
\begin{eqnarray}
ds^2&=&-dt^2 + \frac{a^2(t)}{1-kr^2}dr^2 + R^2 d\Omega_{2}^2
\nonumber
\\
&=&h_{ab}dx^a dx^b + R^2 d\Omega_{2}^2,
\end{eqnarray}
where $R=ar$ is the area radius, $h_{ab}=diag(-1,\frac{a^2}{1-kr^2})$ is the metric of 2-space $(x^0=t,x^1=r)$ and $k=0,\pm1$ denotes the
curvature scalar.

The suface gravity \cite{r37},
\begin{equation}
 \kappa =\frac{1}{2\sqrt{-h}}\partial_a(\sqrt{-h}h^{ab}\partial_b R),
\end{equation}
for any horizon (with area radius $R_h$) in FRW model, can be written as
\begin{equation}
 \kappa = -\left(\frac{R_h}{R_A}\right)^2\left(\frac{1-\frac{\dot{R_A}}{2HR_A}}{R_h}\right)
\end{equation}
i.e.,
\begin{eqnarray}
 \kappa_h&=&-\left(\frac{R_h}{R_A^2}\right)(1-\epsilon), \textrm{for any horizon,}
\nonumber
\\
\textrm{and}~~\kappa_A&=&-\frac{(1-\epsilon)}{R_A}, \textrm{for apparent horizon,}
\end{eqnarray}
with $\epsilon= \frac{\dot{R_A}}{2HR_A}$.

Using this form of surface gravity the extended Hawking temperature is defined as \cite{r38}
\begin{equation}
T_{EH}^{h}=\frac{|\kappa_h|}{2\pi}
\end{equation}
According to Hayward \cite{r28,r29,r30} the unified first law can be expressed as
\begin{equation}
dE=A\psi+WdV
\end{equation}
where $E=\frac{R}{2G}(1-h^{ab}\partial_aR\partial_bR)$ is the total energy inside a sphere of radius R and is termed as Misner-Sharp energy \cite{r28,r29,r30,r35,r39}. Also the energy flux $\psi$ is termed as energy supply vector and W is the work function and are defined as
\begin{equation}
\psi_a=T_a^b \partial_b r+W\partial_a r,
\end{equation}

\begin{equation}
W=-\frac{1}{2}trace T,
\end{equation}
where $T_{ab}$ is the energy momentum tensor.

Further to have a complete thermodynamical study one has to examine the validity of the generalized second law of thermodynamics (GSLT) and thermodynamical equilibrium (TE) on the horizons. For their validity we must have the following inequalities \cite{r40,r41}\\

\begin{equation}
\frac{\partial}{\partial t}S_{TH} \geq 0 ~(for~GSLT)
\end{equation}
and
\begin{equation}
\frac{\partial^2}{\partial t^2}S_{TH} <0 ~(for~ TE)
\end{equation}
where $S_{TH}=S_h+S_{fh}$, with $S_h$ and $S_{fh}$ as the horizon entropy and the entropy of the fluid bounded by the horizon respectively. To obtain fluid entropy $S_{fh}$ one uses Gibb's relation \cite{r42,r43,r44,r45,r46,r47,r48}
$$T_f dS_{fh}=dE_f+pdV_h$$
where $E_f=(\rho V_h)$, is the energy flow across the horizon, $V_h=\frac{4}{3}\pi R_h^3$ is the volume of the fluid ($\rho$, p ) are the energy density and thermodynamic pressure of the fluid and $T_f$ is the temperature of the fluid which is assumed to be same as the extended Hawking temperature on the horizon.As a result the time variation of the fluid entropy is given by
\begin{equation}
\dot{S}_{fh}=\frac{4\pi R_h^2}{T^h_{EH}}(\rho+p)\{\dot{R}_h-HR_h\}
\end{equation}
It should be noted that in deriving the above relation we have used the energy conservation relation for the fluid i.e.,
\begin{equation}
\dot{\rho}+3H(\rho+p)=0
\end{equation}
In the rest of the paper we shall work with units where $8\pi=1=G$. \\\\
{\bf RSII brane world}\\\\
In a flat, homogeneous and isotropic FRW brane extended in 5DAds bulk, the equivalent Friedmann equations [ without dark radiation term ] are given by (\cite{r34,r49}),
\begin{eqnarray}
 H^2&=&\frac{\rho_t}{3},
\nonumber
\\
\dot{H}&=&-\frac{1}{2}(\rho_t+p_t),
\end{eqnarray}
where $\rho_t=\rho+\rho_e$, $p_t=p+p_e$ and
the effective energy density $\rho_e$ and the effective pressure $p_e$, due to embedding of the brane to the bulk have the expressions,
\begin{eqnarray}
 \rho_e&=&\frac{\kappa_5^4 \rho^2}{12},
\nonumber
\\
\rho_e+p_e&=&\frac{\kappa_5^4 \rho(\rho+p)}{9},
\end{eqnarray}
 so that we have,
\begin{equation}
 \frac{\partial}{\partial t}(\rho_e+p_e)=\frac{\kappa_5^4 \rho^2}{9}(f_A-4Hv_A)
\end{equation}
where $\kappa_5$ is 5 dimensional gravitational coupling constant and is related to brane tension ($\lambda $) and 4 dimensional gravitational coupling constant by the relation $\kappa_5^4=\frac{\kappa_4^2 \lambda}{6}$, $v_A~(=\dot{R}_A)$ is the velocity of the apparent horizon and $f_A~(=\dot{v_A})$ is the acceleration of the apparent horizon.

Due to the energy conservation relation for matter (i.e., equation (12)) and from the Bianchi identity we obtain
\begin{equation}
  \dot{\rho_t}+3H(\rho_t+p_t)=0.
\end{equation}
As a result the effective pressure and effective energy density also satisfy the conservation relation
\begin{equation}
 \dot{\rho_e}+3H(\rho_e+p_e)=0.
\end{equation}
Here the work density term can be break up in the following form:
\begin{center}
 $W=W_m+W_e$,
\end{center}
with
\begin{eqnarray}
W_m&=&\frac{\rho-p}{2}
\nonumber
\\
W_e&=&-\frac{\kappa_5^4 \rho p}{12}.
\end{eqnarray}
Also, the energy supply vector can be decomposed as
\begin{center}
 $\psi=\psi_m+\psi_e$
\end{center}
with
\begin{equation}
 \psi_m=-\frac{1}{2}(\rho +p)HRdt+\frac{1}{2}(\rho +p)adr,
\end{equation}
and
\begin{equation}
 \psi_e=-\frac{\kappa_5^4}{12}\rho (\rho+p)HRdt+\frac{\kappa_5^4}{12}\rho (\rho+p)adr.
\end{equation}
As light rays move along the radial direction {\it i.e.,} normal to the surface of the event horizon and we have
$\partial \xi^\pm=\partial_t\mp a\partial_r$ as one form along the normal direction, so $\partial_\pm=-\sqrt{2}(\partial_t\mp \frac{1}{a}\partial_r)$
may be chosen along the tangential direction to the surface of the event horizon. Thus we choose \cite{r38}
\begin{equation}
 \xi_E=\partial_t-\frac{1}{a}\partial_r,
\end{equation}
as the tangential vector to the surface of the event horizon.

Now, projecting the unified first law along $\xi_E$, the first law of thermodynamics of the event horizon is obtained as \cite{r32,r33,r34}
\begin{equation}
 \langle dE,\xi_E \rangle = \kappa_E\langle dA, \xi_E \rangle + \langle WdV, \xi_E \rangle .
\end{equation}
Note that the pure matter energy supply $A\psi_m$, when projected on the event horizon gives the heat flow $\delta Q$ in the Clausius relation
$\delta Q=TdS$. Hence from equation $(22)$, we have
\begin{equation}
 \delta Q=\langle A\psi_m,\xi_E \rangle = \kappa_E\langle dA, \xi_E \rangle - \langle A\psi_e,\xi_E \rangle .
\end{equation}
Using equations $(14), (15), (19) ~and~ (20)$ we obtain (after a simple algebra),
\begin{equation}
 \langle A\psi_m, \xi_E \rangle = \kappa_E R_E \dot{R_E}+\frac{A(HR_E+1)\rho (\rho +p)\kappa_5^4}{12}.
\end{equation}
Now using the Extended Hawking temperature on event horizon, the above equation can be written as
\begin{equation}
 \langle A\psi_m, \xi_E \rangle = T \left\langle \frac{R_E dR_E}{4}-\frac{ \kappa_5^4}{96}\frac{R_A^2 R_E}{1-\epsilon}(HR_E+1)\rho (\rho +p)dt, \xi_E \right\rangle
\end{equation}
Thus comparing with Clausius relation $\delta Q=TdS$ and integrating we have the entropy on the event horizon,
\begin{equation}
 S_E=\frac{A_E}{4}-\frac{ \kappa_5^4}{96}\int \left(\frac{R_A^2 R_E}{1-\epsilon}\right)\left(\frac{HR_E+1}{HR_E-1}\right)\rho (\rho+p)dR_E.
\end{equation}
Similarly for apparent horizon, considering \cite{r32}
\begin{equation}
 \xi_A=\partial_t-(1-2\epsilon)Hr\partial_r,
\end{equation}
as the tangent vector to the surface of the apparent horizon , the expression for entropy becomes
\begin{equation}
 S_A=\frac{A_A}{4}-\frac{ \kappa_5^4}{96}\int \frac{R_A^3}{\epsilon}\rho (\rho +p)dR_A.
\end{equation}

We shall now examine the thermodynamic inequalities (9) and (10) for (a) Perfect fluid and (b) Modified Chaplygin gas\\
(a)\underline{Perfect fluid:}\\
The equation of state parameter $\omega (<-\frac{1}{3})$ is assumed to be constant. For this simple fluid, the horizon entropies take the forms\\
\begin{equation}
 S_E=\frac{A_E}{4}-\frac{\kappa_5^4 \rho_0^2 (\omega +1)}{96}\int \left(\frac{R_A^2 R_E}{1-\epsilon}\right)\left(\frac{HR_E+1}{HR_E-1}\right)\left(\frac{1}{a^{3(\omega+1)}}\right)^2dR_E.
\end{equation}
and
\begin{equation}
 S_A=\frac{A_A}{4}-\frac{\kappa_5^4 \rho_0^2(\omega +1)}{96}\int \frac{R_A^3}{\epsilon}\left(\frac{1}{a^{3(\omega+1)}}\right)^2dR_A.
\end{equation}
where $\rho=\frac{\rho_0}{a^{3(\omega +1)}}$, and $\rho_0$ is an arbitrary constant.\\

Hence the time derivative of the total entropy (i.e., entropy of the horizon + entropy of the fluid) are given by
\begin{equation}
\dot{S}_{TA}=\frac{R_A v_A}{4}-\frac{R_A^3}{8}(\rho_e+p_e)+\frac{R_A v_A(v_A-1)}{2(2-v_A)},~\textrm{for apparent horizon}
\end{equation}
\begin{equation}
\dot{S}_{TE}=\frac{R_Ev_E}{4}-\frac{R_A^2R_E}{2(2-v_A)}\{\frac{v_E+2}{4}(\rho_e+p_e)+v_AH^2\},~~\textrm{for event horizon}
\end{equation}
Again, taking the time derivative of $\dot{S}_{TA}$ and $\dot{S}_{TE}$ we have\\
\begin{equation}
\ddot{S}_{TA}=\frac{R_Af_A}{4}\{1-\frac{2}{(v_A-2)^2}(v_A^2-4v_A+2)\}-\frac{R_A^2}{8}\{3v_A(\rho_e+p_e)+8R_A\frac{\partial (\rho_e+p_e)}{\partial t}\}-\frac{v_A^2(v_A-1)}{2(v_A-2)},~~\textrm{for apparent horizon}
\end{equation}
$$\ddot{S}_{TE}=\frac{R_Ef_E}{4}\{1-\frac{R_A^2(\rho_e+p_e)}{2(2-v_A)}\}+\frac{v_E^2}{4}-\frac{R_A^2R_E}{8(2-v_A)}\{(v_E+2)\left(2\frac{v_A}{R_A}+\frac{v_E}{R_E}+\frac{f_A}{2-v_A}\right)(\rho_e+p_e)+\frac{\partial (\rho_e+p_e)}{\partial t}\}$$
\begin{equation}
+4v_AH^2\{\frac{v_E}{R_E}+\frac{2f_A}{v_A(2-v_A)}\}~~\textrm{for event horizon}.
\end{equation}
In the above equations (31)-(34), $\rho_e+p_e=\frac{\kappa_5^4}{9}(1+\omega)\rho^2$ and $\frac{\partial}{\partial t}(\rho_e+p_e)=\frac{\kappa_5^4}{9}(f_A-4Hv_A)\rho^2$, $v_E (=\dot{R}_E)$ and $f_E (=\dot{v}_E)$ are respectively the velocity and the acceleration of the event horizon.
Due to complicated expressions we can not infer the sign of the above expressions, so we examine the validity of GSLT and TE graphically in figures 1 and 2 for H=1, $\kappa_5=1~and~R_E=3$. Also restrictions on $\omega$ to satisfy the inequalities (9), (10) are shown in table-1.\\

\begin{figure}
\begin{minipage}{0.4\textwidth}
\includegraphics[width=1.0\linewidth]{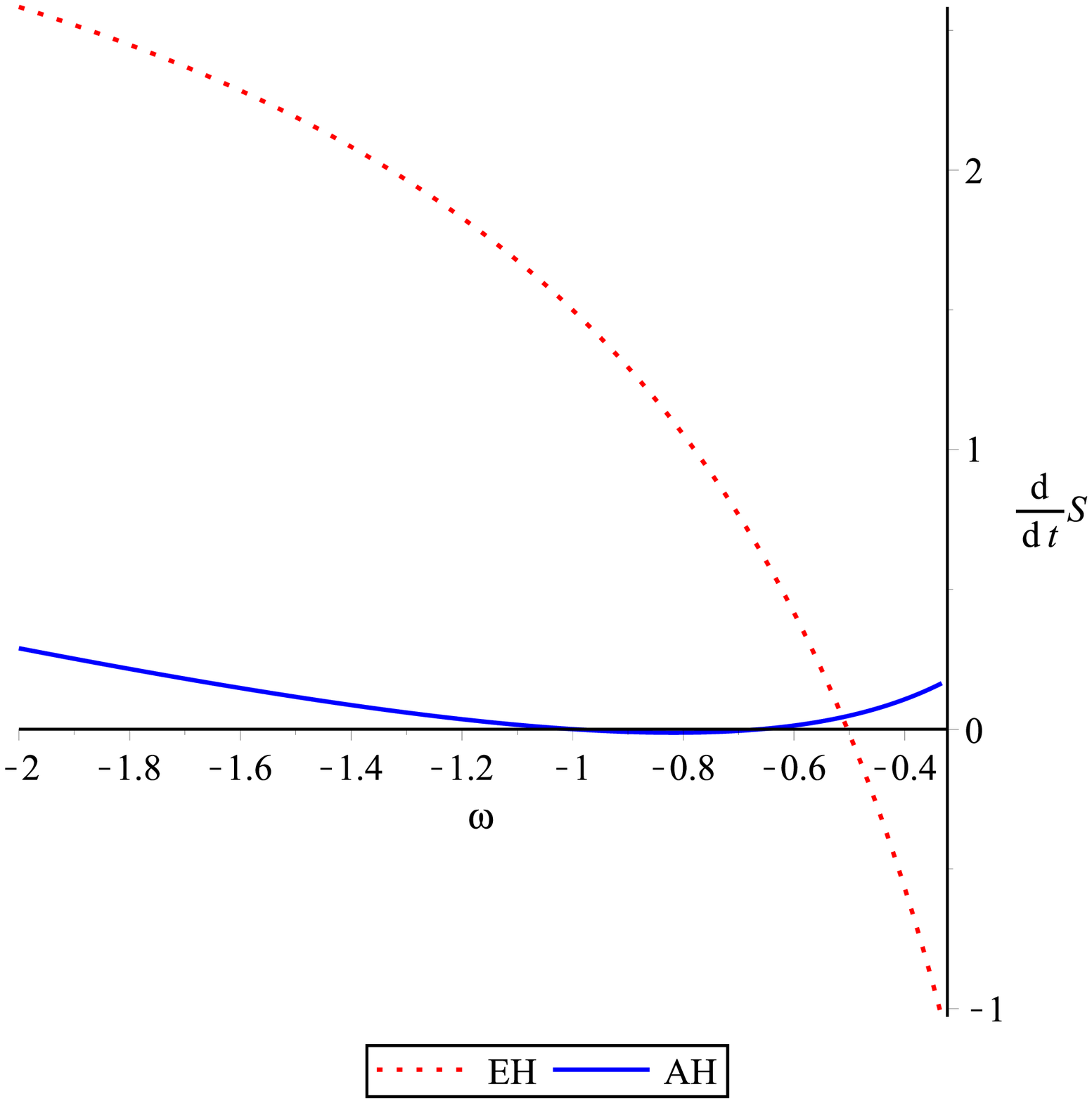}
Fig-1:GSLT for RSII brane for perfect fluid with constant equation of state
\end{minipage}
\begin{minipage}{0.4\textwidth}
\includegraphics[width=1.0\linewidth]{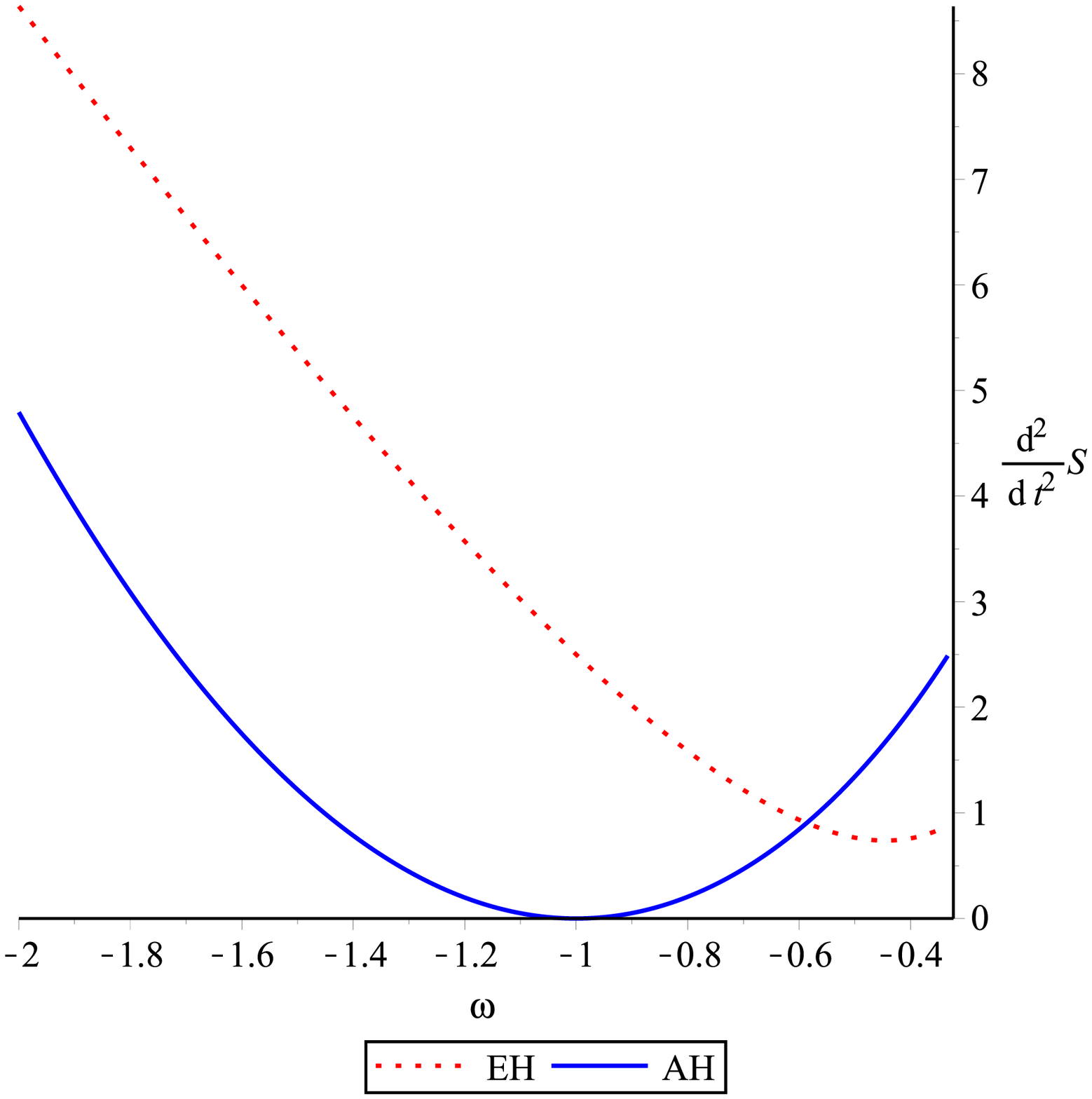}
Fig-2:TE for RSII brane for perfect fluid with constant equation of state
\end{minipage}
\end{figure}

(b)\underline{ Modified Chaplygin Gas:}\\
The equation of state for Modified Chaplygin Gas is written as \cite{r41,r50},
\begin{equation}
 p=\gamma \rho -\frac{B}{\rho^n},
\end{equation}
where $\gamma (\leq 1)$ and B, n are positive constants. Now solving the energy conservation equation (17), we have
\begin{equation}
 \rho^{n+1}=\frac{1}{\gamma +1}\left[B+\left(\frac{C}{a^3}\right)^{(\gamma+1)(n+1)}\right],
\end{equation}
where C is an arbitrary constant.

In this model the velocity of the apparent horizon is $v_A=\frac{3C(1+\gamma)}{2(Ba^\mu +C)}$, $\mu=3(1+n)(1+\gamma)$. The radius of the event horizon can be expressed in terms of hypergeometric function as $$R_E=R_{1~~2}F_1\left[\frac{1}{2(n+1)}, \frac{1}{\mu}, 1+\frac{1}{\mu}, \frac{-C}{Ba^\mu}\right],$$ where $R_1=\frac{\sqrt{3}(1+\gamma)^{\frac{1}{2(n+1)}}}{B^{\frac{1}{2(n+1)}}}$.\\
Hence expressions for horizon entropy become:\\
\begin{eqnarray}
 S_E=\frac{A_E}{4}-\frac{ \kappa_5^4}{96 (\gamma+1)^{\frac{2}{n+1}}}\int \left(\frac{R_A^2 R_E}{1-\epsilon}\right)\left(\frac{HR_E+1}{HR_E-1}\right)\left[B+\left(\frac{C}{a^3}\right)^{(\gamma+1)(n+1)}\right]^{\frac{2}{n+1}} \times
\nonumber
\\
\left[\gamma+1-B(\gamma+1)\left\lbrace B+\left(\frac{C}{a^3}\right)^{(\gamma+1)(n+1)}\right\rbrace^{-1}\right]dR_E,
\end{eqnarray}
and
\begin{equation}
 S_A=\frac{A_A}{4}-\frac{\kappa_5^4 }{96 (\gamma+1)^{\frac{2}{n+1}}}\int \frac{R_A^3}{\epsilon}\left[B+\left(\frac{C}{a^3}\right)^{(\gamma+1)(n+1)}\right]^{\frac{2}{n+1}}\left[\gamma+1-B(\gamma+1)\left\lbrace B+\left(\frac{C}{a^3}\right)^{(\gamma+1)(n+1)}\right\rbrace^{-1}\right]dR_A,
\end{equation}
Now, the time derivative of the total entropy are given by\\
\begin{equation}
\dot{S}_{TA}=\frac{R_A v_A}{4}-\frac{R_A^3}{8}(\rho_e+p_e)+\frac{R_A v_A(v_A-1)}{2(2-v_A)},~\textrm{for apparent horizon}
\end{equation}
\begin{equation}
\dot{S}_{TE}=\frac{R_Ev_E}{4}-\frac{R_A^2R_E}{2(2-v_A)}\{\frac{v_E+2}{4}(\rho_e+p_e)+v_AH^2\},~~\textrm{for event horizon}
\end{equation}
Again, taking the derivative of $\dot{S}_{TA}$ and $\dot{S}_{TE}$ we have\\
\begin{equation}
\ddot{S}_{TA}=\frac{R_Af_A}{4}\{1-\frac{2}{(v_A-2)^2}(v_A^2-4v_A+2)\}-\frac{R_A^2}{8}\{3v_A(\rho_e+p_e)+8R_A\frac{\partial (\rho_e+p_e)}{\partial t}\}-\frac{v_A^2(v_A-1)}{2(v_A-2)},~~\textrm{for apparent horizon}
\end{equation}
$$\ddot{S}_{TE}=\frac{R_Ef_E}{4}\{1-\frac{R_A^2(\rho_e+p_e)}{2(2-v_A)}\}+\frac{v_E^2}{4}-\frac{R_A^2R_E}{8(2-v_A)}\{(v_E+2)\left(2\frac{v_A}{R_A}+\frac{v_E}{R_E}+\frac{f_A}{2-v_A}\right)(\rho_e+p_e)+\frac{\partial (\rho_e+p_e)}{\partial t}\}$$
\begin{equation}
+4v_AH^2\{\frac{v_E}{R_E}+\frac{2f_A}{v_A(2-v_A)}\}~~\textrm{for event horizon}.
\end{equation}
In the above equations (39)-(42), $\rho_e+p_e=\frac{\kappa_5^4}{9}\rho^2(1+\gamma-\frac{B}{\rho^{n+1}})$ , $\frac{\partial}{\partial t}(\rho_e+p_e)=\frac{\kappa_5^4}{9}(f_A-4Hv_A)\rho^2$, where $\rho$ is given by equation (36).

As before, due to complicated form of the above entropy variations, the thermodynamical inequalities (9) and (10) are examined graphically in figure 3 and 4. In particular $\dot{S}_{TA}, \ddot{S}_{TA}, \dot{S}_{TE}, \ddot{S}_{TE}$ are plotted against $\gamma$  in the figures considering H=1 , $\kappa_5=1,~a=1,~n=0.25,~B=2~and~C=1$. Also in table-2 bounds on $\gamma$ are shown to satisfy the inequalities (9) and (10).\\
\begin{figure}
\begin{minipage}{0.4\textwidth}
\includegraphics[width=1.0\linewidth]{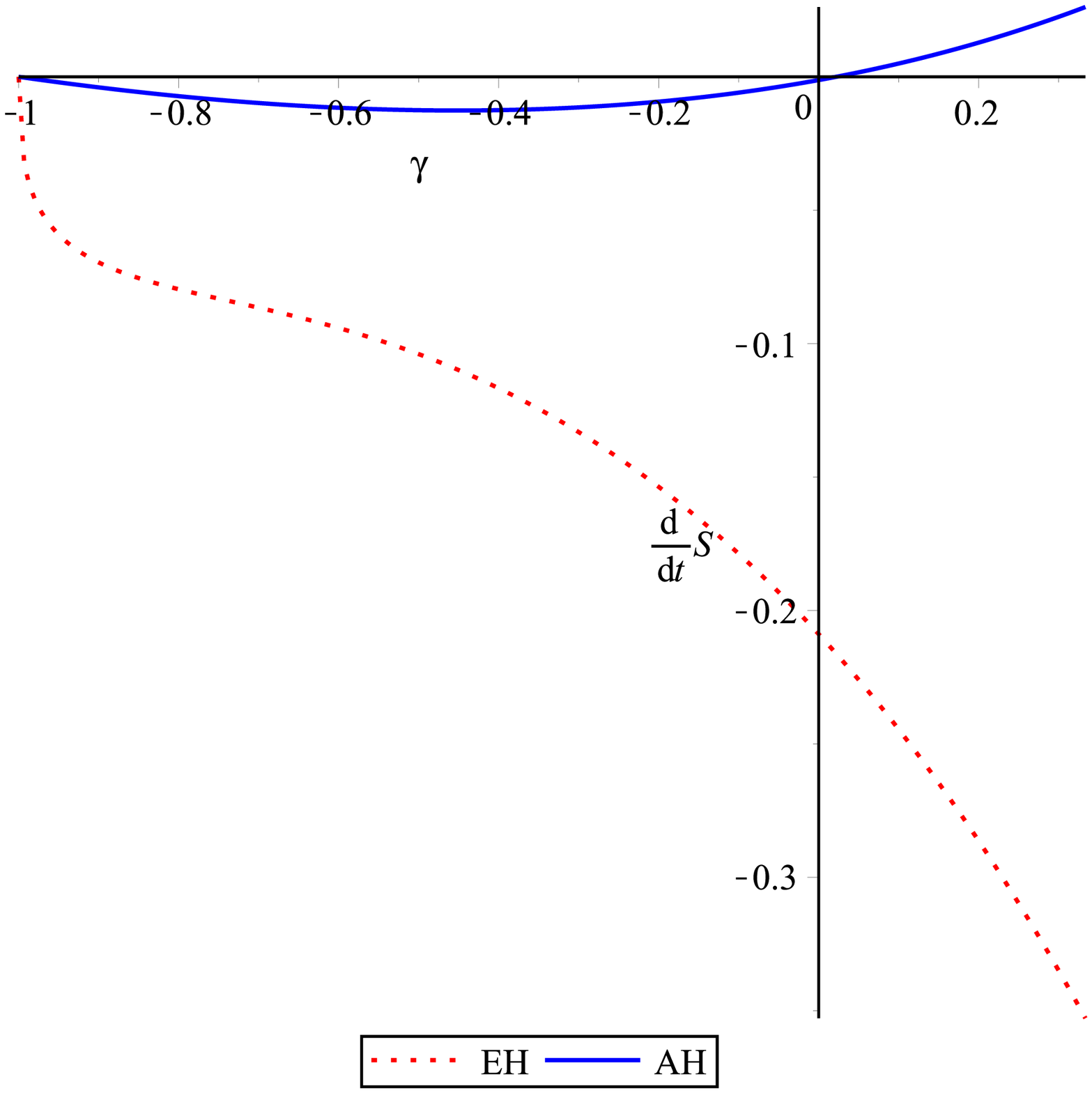}
Fig-3:GSLT for RSII brane for Modified Chaplygin Gas
\end{minipage}
\begin{minipage}{0.4\textwidth}
\includegraphics[width=1.0\linewidth]{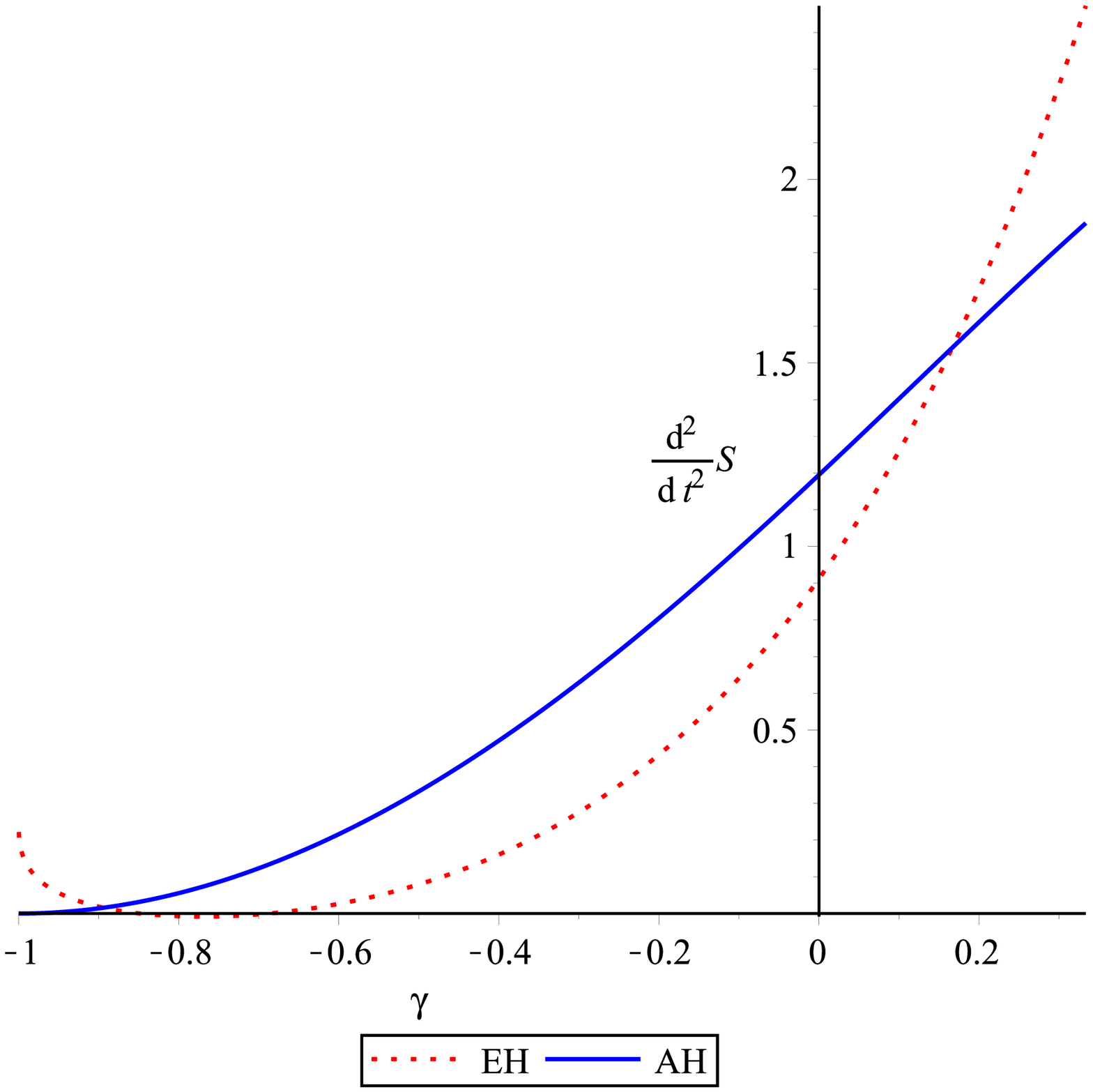}
Fig-4:TE for RSII brane for Modified Chaplygin Gas
\end{minipage}
\end{figure}

{\bf DGP brane world}\\
In a flat, homogeneous and isotropic brane, the Friedmann equation in DGP model is given by \cite{r51,r52}
\begin{equation}
 H^2-\tilde{\epsilon}\frac{H}{r_c}=\frac{\rho}{3}
\end{equation}
where $r_c$ is the crossover scale which determines the transition from 4D to 5D behavior and $\tilde{\epsilon}=\pm1$ corresponds to
standard DGP(+) model (self accelerating without any form of dark energy) and DGP(-) model (not self accelerating, requires dark energy)
respectively.

From (43) and using the conservation equation (17), it can be shown that
\begin{equation}
 \dot{H}=-\frac{1}{2}\left[\rho+p +\frac{\tilde{\epsilon} (\rho +p)}{2Hr_c-\tilde{\epsilon}}\right].
\end{equation}
Thus we have
\begin{eqnarray}
 \rho_e+p_e&=&\frac{\tilde{\epsilon}(\rho+p)}{2Hr_c-\tilde{\epsilon}},
 \nonumber
 \\
 \frac{\partial (\rho_e+p_e)}{\partial t}&=&\frac{2\rho \tilde{\epsilon}}{3(2Hr_c-\tilde{\epsilon})}\left(f_A-2Hv_A^2+\frac{2H^2v_A^2r_c}{2Hr_c-\tilde{\epsilon}}\right).
\end{eqnarray}

Considering $\xi$ as given by (27) (for apparent horizon) and by equation (21) (for event horizon) and proceeding in the same way as before, the expressions of entropy on the horizon (apparent / event) for
the validity of the unified first law are given by
\begin{equation}
 S_A=\frac{A_A}{4}-\frac{1}{16}\tilde{\epsilon} \int \left(\frac{R_A^3}{\epsilon}\right)\left(\frac{\rho+p}{2Hr_c-\tilde{\epsilon}}\right)dR_A,
\end{equation}
and
\begin{equation}
 S_E=\frac{A_E}{4}-\frac{1}{16}\tilde{\epsilon}\int \left(\frac{R_A^2 R_E}{1-\epsilon}\right)\left(\frac{HR_E+1}{HR_E-1}\right)\frac{\rho+p}{2Hr_c-\tilde{\epsilon}}dR_E,
\end{equation}

As in RSII brane we take the time derivative of fluid entropy from Gibb's equation and combining with the time derivative of the horizon entropy expressions, we have\\

\begin{equation}
\dot{S}_{TA}=\frac{R_A v_A}{4}-\frac{R_A^3}{8}(\rho_e+p_e)+\frac{R_A v_A(v_A-1)}{2(2-v_A)},~\textrm{for apparent horizon}
\end{equation}
and
\begin{equation}
\dot{S}_{TE}=\frac{R_Ev_E}{4}-\frac{R_A^2R_E}{2(2-v_A)}\{\frac{v_E+2}{4}(\rho_e+p_e)+v_AH^2\},~\textrm{for event horizon}
\end{equation}
Again, taking the time derivative of $\dot{S}_{TA}$ and $\dot{S}_{TE}$ we have\\
\begin{equation}
\ddot{S}_{TA}=\frac{R_Af_A}{4}\{1-\frac{2}{(v_A-2)^2}(v_A^2-4v_A+2)\}-\frac{R_A^2}{8}\{3v_A(\rho_e+p_e)+8R_A\frac{\partial (\rho_e+p_e)}{\partial t}\}-\frac{v_A^2(v_A-1)}{2(v_A-2)},~\textrm{for apparent horizon}
\end{equation}
$$\ddot{S}_{TE}=\frac{R_Ef_E}{4}\{1-\frac{R_A^2(\rho_e+p_e)}{2(2-v_A)}\}+\frac{v_E^2}{4}-\frac{R_A^2R_E}{8(2-v_A)}\{(v_E+2)\left(2\frac{v_A}{R_A}+\frac{v_E}{R_E}+\frac{f_A}{2-v_A}\right)(\rho_e+p_e)+\frac{\partial (\rho_e+p_e)}{\partial t}\}$$
\begin{equation}
+4v_AH^2\{\frac{v_E}{R_E}+\frac{2f_A}{v_A(2-v_A)}\} ~\textrm{for event horizon}.
\end{equation}
In the above equations (48)-(51), $\rho_e+p_e$ and $\frac{\partial}{\partial t}(\rho_e+p_e)$ are to be substituted from equation (45). \\
For perfect fluid with constant equation of state, the expressions of horizon entropy become:
\begin{equation}
  S_A=\frac{A_A}{4}-\frac{1}{16}\tilde{\epsilon}(\omega+1)\rho_0 \int \left(\frac{R_A^3}{\epsilon}\right)\left(\frac{1}{2Hr_c-\tilde{\epsilon}}\right)\frac{1}{a^{3(\omega+1)}}dR_A,
\end{equation}
and
\begin{equation}
 S_E=\frac{A_E}{4}-\frac{1}{16}\tilde{\epsilon}(\omega+1)\rho_0\int \left(\frac{R_A^2 R_E}{1-\epsilon}\right)\left(\frac{HR_E+1}{HR_E-1}\right)\left(\frac{1}{2Hr_c-\tilde{\epsilon}}\right)\frac{1}{a^{3(\omega+1)}}dR_E,
\end{equation}

Similar to RSII model, $\dot{S}_{TA}, \ddot{S}_{TA}, \dot{S}_{TE}, \ddot{S}_{TE}$ have been plotted against $\omega$ for apparent horizon (AH) and event horizon (EH) in figures 5 and 6 respectively, considering H=1.5 , $\epsilon=1,~R_E=3$ and $r_c=2$.\\
\begin{figure}
\begin{minipage}{0.4\textwidth}
\includegraphics[width=1.0\linewidth]{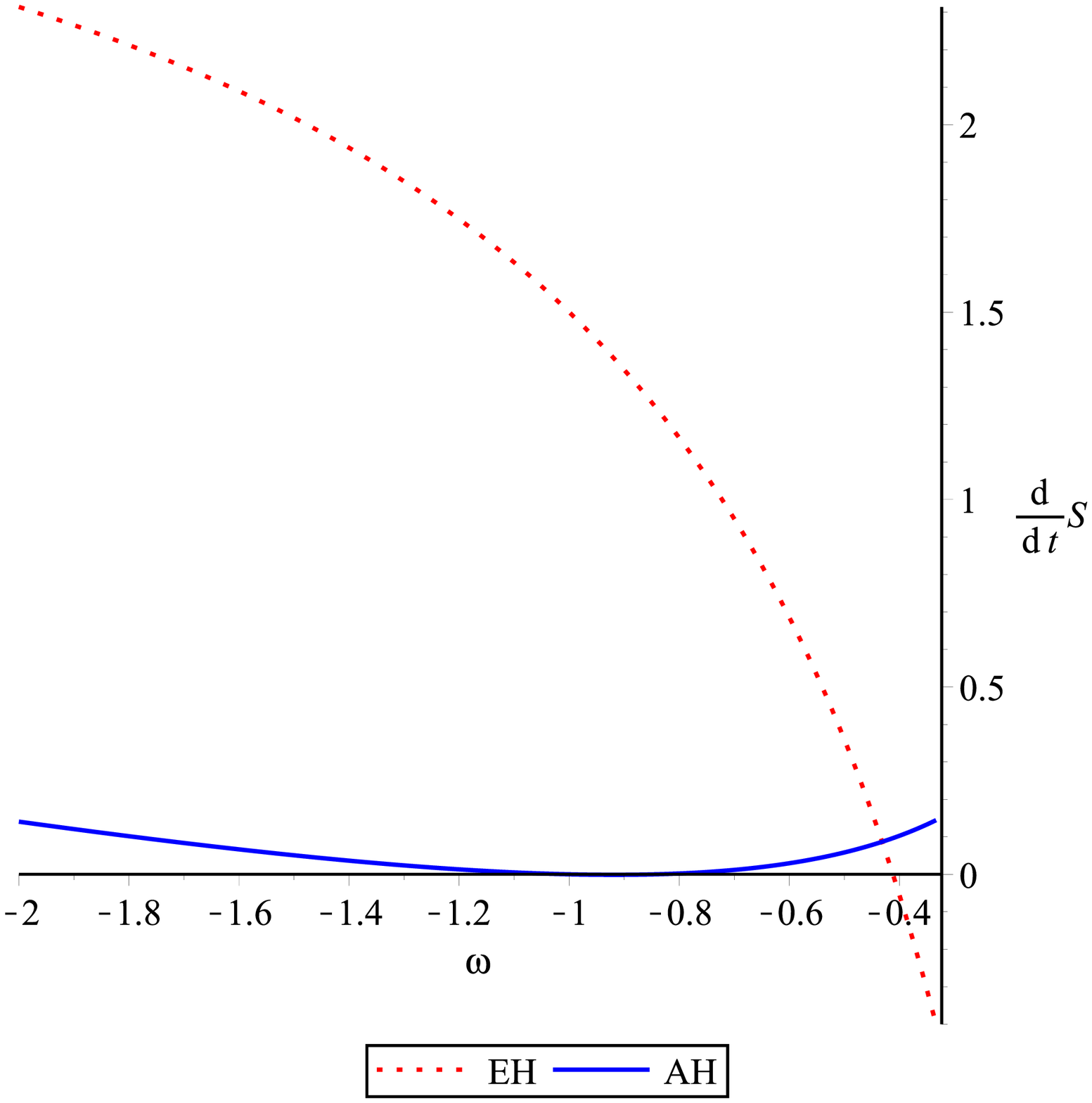}
Fig-5:GSLT for DGP brane for perfect fluid with constant equation of state
\end{minipage}
\begin{minipage}{0.4\textwidth}
\includegraphics[width=1.0\linewidth]{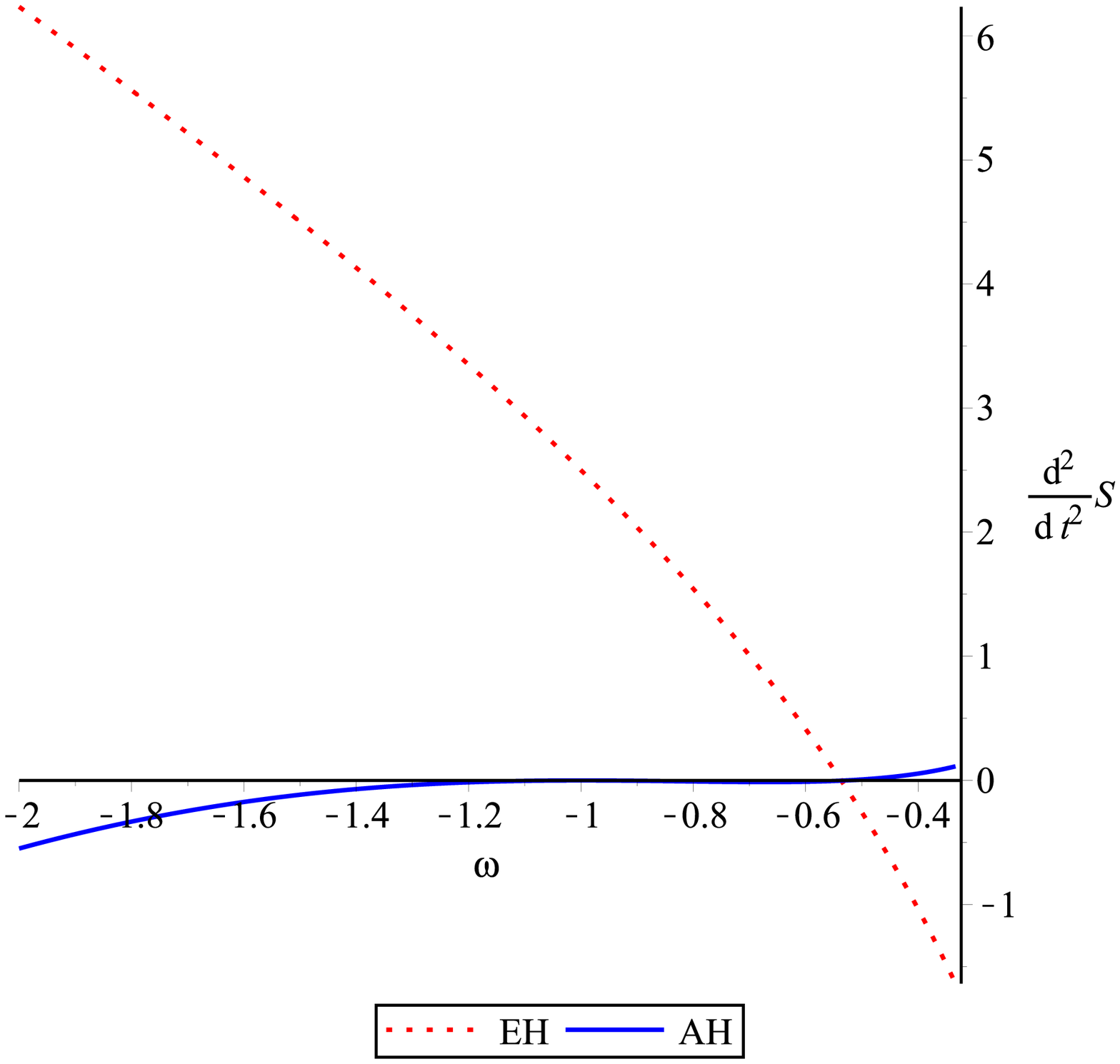}
Fig-6:TE for DGP brane for perfect fluid with constant equation of state
\end{minipage}
\end{figure}

For Modified Chaplygin gas, the horizon entropy takes the form
\begin{eqnarray}
 S_A=\frac{A_A}{4}-\frac{1}{16(\gamma+1)^{\frac{1}{n+1}}} \tilde{\epsilon}\int \left(\frac{R_A^3}{\epsilon}\right)\left(\frac{1}{2Hr_c-\tilde{\epsilon}}\right)\left[B+\left(\frac{C}{a^3}\right)^{(\gamma+1)(n+1)}\right]^{\frac{1}{n+1}} \times
\nonumber
\\
\left[\gamma+1-B(\gamma+1)\left\lbrace B+\left(\frac{C}{a^3}\right)^{(\gamma+1)(n+1)}\right\rbrace^{-1}\right]dR_A,
\end{eqnarray}
and
\begin{eqnarray}
 S_E=\frac{A_E}{4}-\frac{1}{16(\gamma+1)^{\frac{1}{n+1}}} \tilde{\epsilon}\int \left(\frac{R_A^2 R_E}{1-\epsilon}\right)\left(\frac{HR_E+1}{HR_E-1}\right)
\left(\frac{1}{2Hr_c-\tilde{\epsilon}}\right)\left[B+\left(\frac{C}{a^3}\right)^{(\gamma+1)(n+1)}\right]^{\frac{1}{n+1}} \times
\nonumber
\\
\left[\gamma+1-B(\gamma+1)\left\lbrace B+\left(\frac{C}{a^3}\right)^{(\gamma+1)(n+1)}\right\rbrace^{-1}\right]dR_E.
\end{eqnarray}

We have plotted $\dot{S}_{TA}, \ddot{S}_{TA}, \dot{S}_{TE}, \ddot{S}_{TE}$  against $\gamma$ for apparent horizon (AH) and event horizon (EH) in figures 7 and 8 respectively, considering H=1.5 , $\epsilon=1,~a=1,~n=0.25,~B=2,~C=1$ and $r_c=2$.\\
\begin{figure}
\begin{minipage}{0.4\textwidth}
\includegraphics[width=1.0\linewidth]{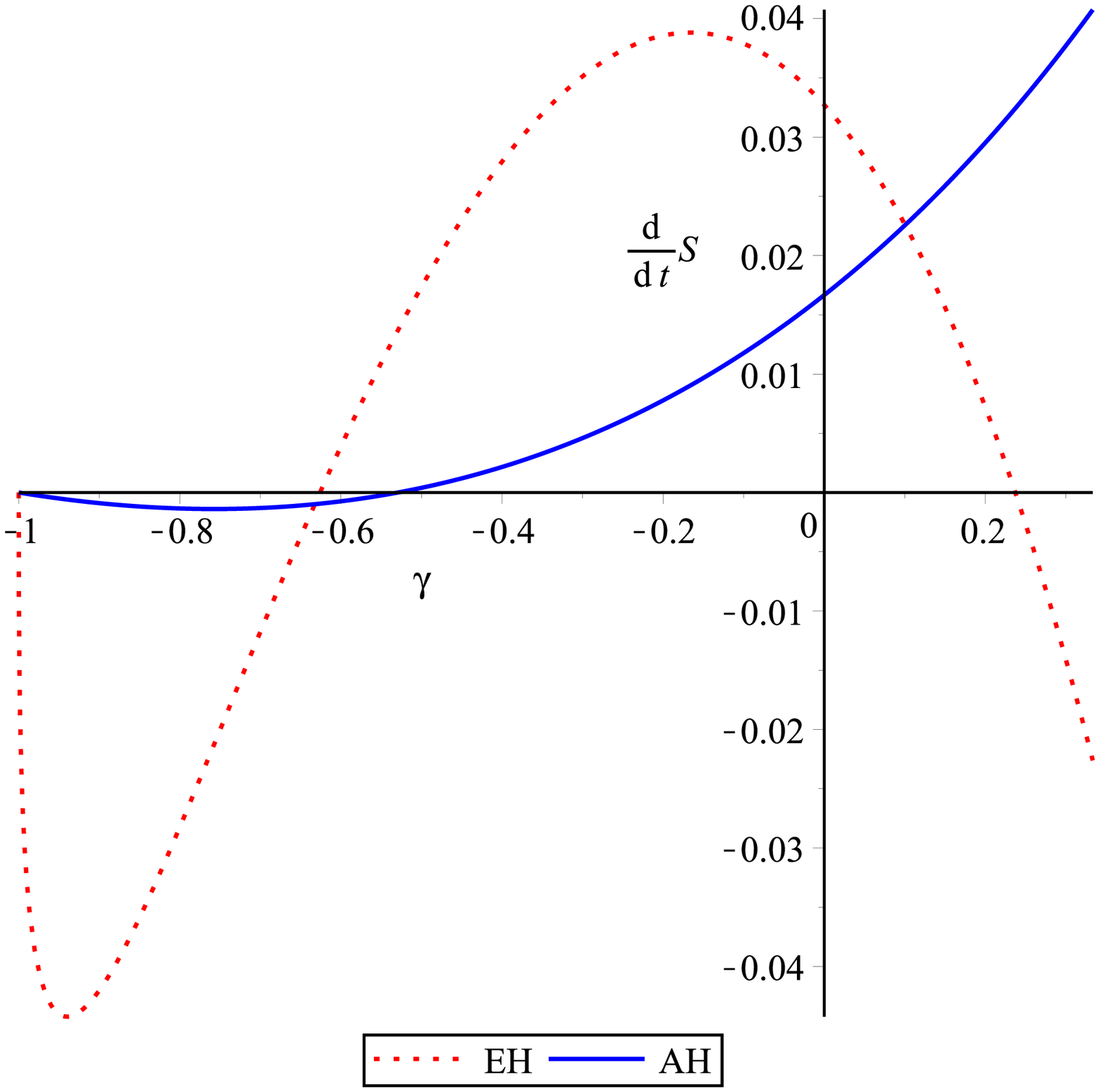}
Fig-7:GSLT for DGP brane for Modified Chaplygin Gas
\end{minipage}
\begin{minipage}{0.4\textwidth}
\includegraphics[width=1.0\linewidth]{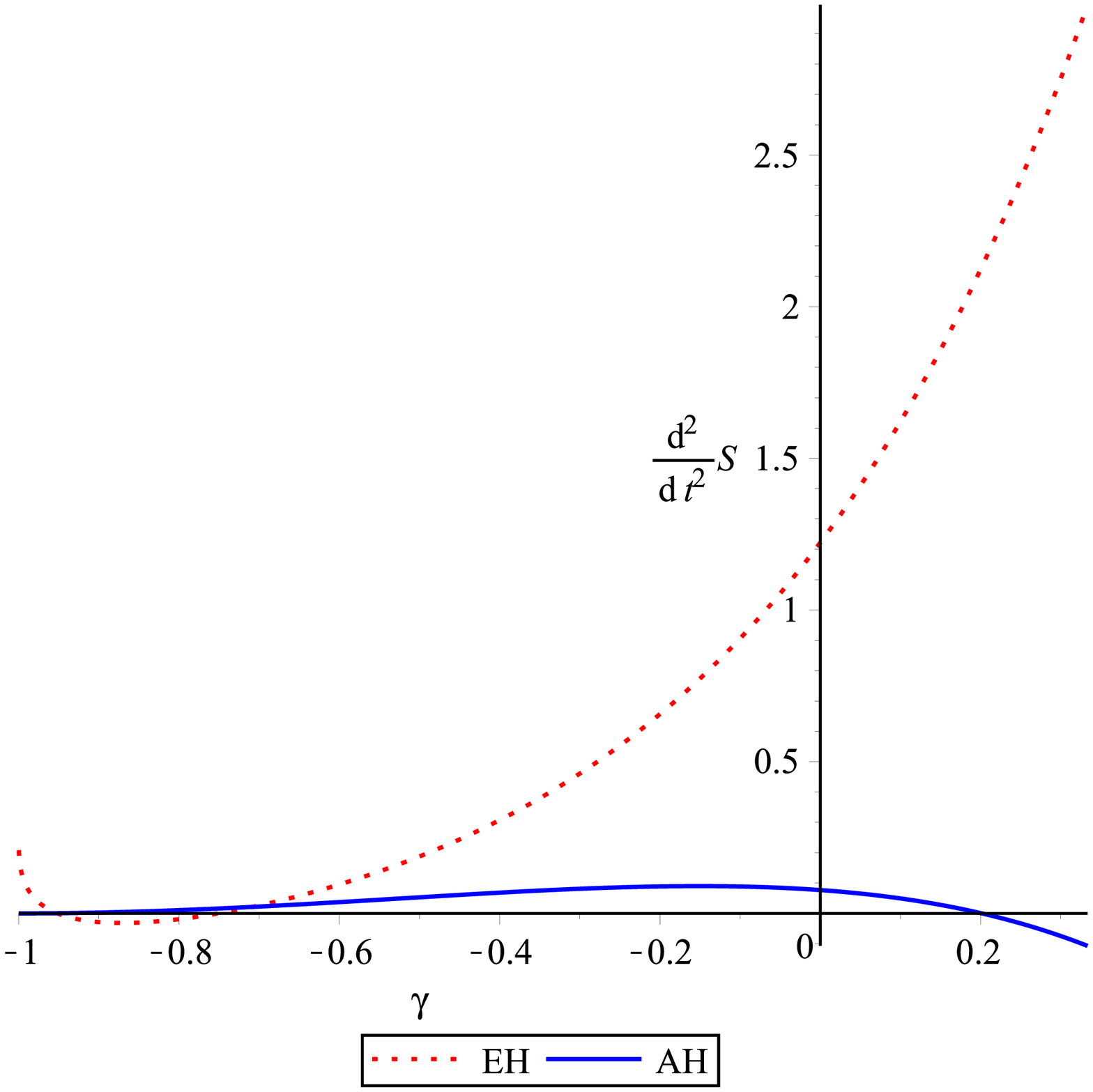}
Fig-8:TE for DGP brane for Modified Chaplygin Gas
\end{minipage}
\end{figure}


Thus in the present work, we have considered Universal thermodynamics for brane world scenario both in RSII model and in DGP model when the FRW Universe is bounded by the horizon (event/apparent). The matter content in the Universe is chosen as one of the following fluids:\\i) perfect fluid with constant equation of state which may be considered as normal fluid or an exotic fluid depending on the equation of state parameter\\ii) modified Chaplygin gas model, a unified model of dark matter and dark energy, extending upto $\Lambda$CDM.\\
The temperature of the fluid as well as that of the horizon is taken as extended Hawking temperature and we have examined the validity of the unified first law on the horizon. It turns out that the entropy on the horizon is no longer the Bekenstein entropy, rather the correction term is in integral form.\\
\begin{center}
Table 1: Conditions for  GSLT and TE to hold when Universe is filled with perfect fluid having constant equation of state
\end{center}
\begin{center}
\begin{tabular}{|c|c|c|}
\hline Horizons in Brane world & GSLT & TE\\
\hline \hline AH for RSII Brane & $\omega \geq -0.64 $ or $\omega \leq -0.98$ & Does not hold\\
\hline EH for RSII Brane & $\omega \leq -0.51$ & Does not hold \\
\hline AH for DGP Brane & $\omega \geq -0.82$ or $\omega \leq -1$ & $\omega<-1.08$ or $-0.89<\omega<-0.54$\\
\hline EH for DGP Brane & $\omega \leq -0.41$ & $\omega > -0.53$\\
\hline
\end{tabular}
\end{center}

\begin{center}
Table 2: Conditions for GSLT and TE to hold when Universe is filled with Modified Chaplygin Gas
\end{center}
\begin{center}
\begin{tabular}{|c|c|c|}
\hline Horizons in Brane world & GSLT & TE\\
\hline \hline AH for RSII Brane & $\gamma \geq 0.02 $ & Does not hold\\
\hline EH for RSII Brane &  Does not hold & $-0.85<\gamma<-0.68$\\
\hline AH for DGP Brane & $\gamma=-1$ or $\gamma \geq -0.529$ & $\gamma> 0.20$\\
\hline EH for DGP Brane & $\gamma=-1$ or $-0.62 \leq \gamma \leq 0.23$ & $-0.95<\gamma <-0.74$\\
\hline
\end{tabular}
\end{center}
From the figures as well as from the tables we see that GSLT holds for both the horizons and for both the fluids with some restrictions on the equation of state parameter $'\omega'$ or on the parameter $'\gamma'$ (for MCG) except that GSLT does not hold at all at the event horizon for RSII brane with MCG model. On the otherhand, thermodynamical equilibrium does not hold at the apparent horizon for RSII brane model for both the fluids while for event horizon in RSII brane model TE does not hold for perfect fluid and it holds for MCG in a restricted range of $'\gamma'$ (see Table 2).

For DGP brane model, TE as well as GSLT holds for both the brane models and for both the fluids in some restricted range of the parameters $\gamma~and~\omega$. So, from the above thermodynamical analysis, we can not conclude that one horizon is more favourable than the other.

From the field equations (13), using (14) we see that the Hubble parameter depends on the brane tension and the effect of the higher dimension in RSII brane model. Similarly, for DGP brane scenario the field equations (43) and (44) depends on the crossover scale ($r_c$) which determines the transition from 4$-$dimensional to 5$-$dimensional behavior. As $R_A \left(=\frac{1}{\sqrt{H^2+\frac{k}{a^2}}}\right)$ and $R_E \left(=a \int_a^\infty \frac{da}{a^2H}\right)$ depend on the Hubble parameter so the radius of horizons depend on the fifth dimensional effect. In the present work, the factor $\kappa_5^4$ in RSII brane model and '$r_c$' in DGP model represents  the effect of the correseponding brane model.

Finally, it should be noted that in earlier works in this direction by Cai et al, \cite{r32,r33,r34}, it was concluded that Universal thermodynamics in these modified gravity theories are not equilibrium in nature and there is entropy production term for the validity of unified first law. On the contrary, in the present work we have shown that Universal thermodynamics remains equilibrium in nature in modified gravity theories but the Bekenstein entropy needs some correction term. Therefore, we conclude that one should consider corrections to Bekenstein entropy so that Universal thermodynamics in modified gravity theories still remain equilibrium in nature.

\section*{Acknowledgement}
The author S.M. is thankful to UGC for NET-JRF.
The author S.S. is thankful to UGC-BSR Programme of Jadavpur University for awarding Research Fellowship.
S.C. is thankful to UGC-DRS programme, Department of Mathematics, J.U.

\end{document}